\documentclass[aps,twocolumn,showpacs]{revtex4}
\usepackage{graphicx}
\begin{document}

\title{Topological defects in 1D elastic waves}

\author{A. Morales}
\email{mori@fis.unam.mx}
\affiliation{Centro de Ciencias F\'{\i}sicas, UNAM, A.P. 48-3, 62251, Cuernavaca, Mor.,
M\'exico }

\author{R. A. M\'endez-S\'anchez}
\affiliation{Centro de Ciencias F\'{\i}sicas, UNAM, A.P. 48-3, 62251, Cuernavaca, Mor.,
M\'exico }

\author{J. Flores}
\altaffiliation{Permanent address: Instituto de F\'{\i}sica,
UNAM, P. O. Box 20-364, 01000 M\'exico, D. F. M\'exico}
\affiliation{Centro de Ciencias F\'{\i}sicas, UNAM, A.P. 48-3, 62251, Cuernavaca, Mor.,
M\'exico }

\begin{abstract}
It has been recently shown theoretically that a topological defect in a
1D periodic potential may give rise to {\it two} localized states
within the energy gaps. In this work we present an experimental realization
of this effect for the case of torsional waves in elastic rods. We
also show numerically that {\it three,} or even more, localized states
can be present if the parameters characterizing the topological defect
are suitably varied.
\end{abstract}
\maketitle

\begin{figure}[tb]
\includegraphics[width=\columnwidth]{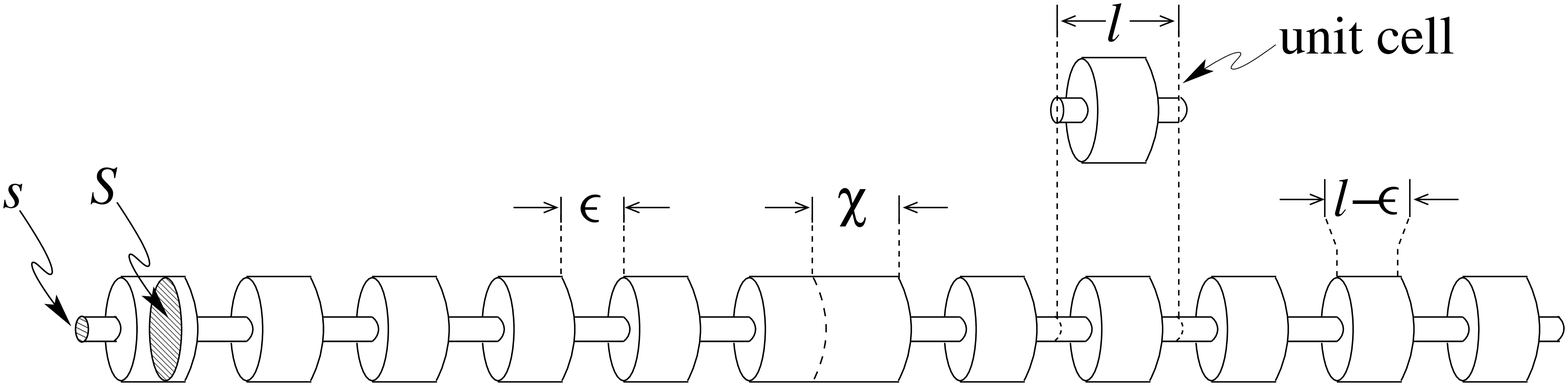}
\caption{Geometry of the aluminum rod.  The length of the unit cell is $l$ and
$\epsilon$ is the width of the notch.
Here $S$ and $s$ are the cross section areas of the rod and notches,
respectively. The defective cell has length $(l-\epsilon)+\chi$.}
\label{rod}
\end{figure}

\section{introduction}

Gumen {\it et al }\cite{Gumenetal} have considered recently the consequences
of introducing a topological defect, that is, a defect that breaks the long range order in an otherwise 1D periodic structure.
They analyze the case of two semi-infinite equal lattices that match
at a point which is not at the center of the unit cell. As an example,
they consider the potential
\begin{equation}
V(x)=V_{0}\cos \left(2\pi \frac{x}{d}-\frac{\Delta }{2}\textrm{sgn}(x)\right),
\end{equation}
where $\Delta $ is related to the strength of the defect. The authors
then solve the stationary Schr\"odinger equation, which is Mathieu's
equation at both sides of the defect, and find that for special values
of the defect strength $\Delta $, two levels appear in the forbidden
band. The situation is different from what occurs with a point defect,
where typically only one level lies in the energy gaps.  In this paper we shall
demonstrate this effect for the normal-mode frequencies of  torsional
waves in rods with notches, both numerically and experimentally.

\section{Torsional waves for rods with topological defects:  theory and experiment}

\begin{figure}
\includegraphics[width=\columnwidth]{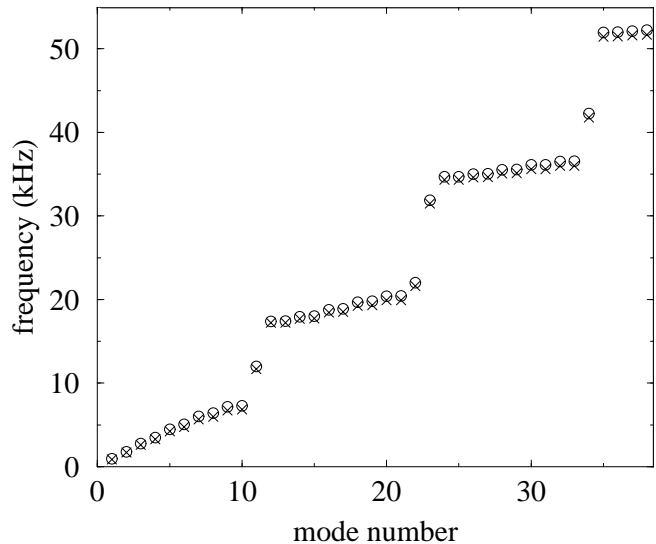}
\caption{Experimental (crosses) and numerical (circles) frequency
spectra of torsional waves for a defective rod with $N=10$,
$l=10.00$~cm, $\epsilon=1.00$~cm, and $\chi=6.00$~cm. In the case
of the fourth band only four frequencies are shown.}
\label{espectro}
\end{figure}

\begin{figure}
\includegraphics[width=\columnwidth]{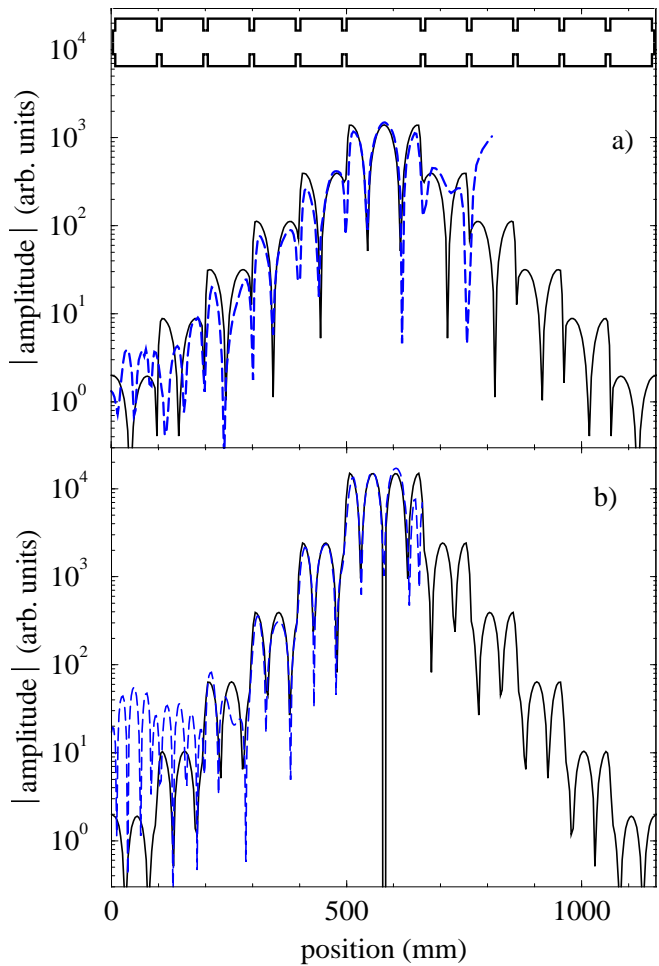}
\caption{Experimental (dashed lines) and theoretical (continuous lines)
wave amplitudes for a) n=22 and b) n=23 nodes. Since the vertical scale is
logarithmic, it can be seen that theory and experiment agree, except for
very small amplitudes. Experimental values at the right-hand side
were not measured, since in this portion of the rod the exciter interferes
with the detector.}
\label{waveamplitudes}
\end{figure}

In this note we deal with torsional vibrations of the elastic
rod shown in Fig.~\ref{rod}. The rods consist of $N$ unit cells plus a defective
cell, which is the topological defect. With the exception of the defect,
each cell is formed by three cylinders, one of length $l-\epsilon $
with cross section area $S$ and two cylinders of length $\epsilon /2$ with
cross section area $s$. In the defect the central cylinder has length
$(l-\epsilon)+\chi $. The wavelength is much larger than the radius
of the rod, so the system is indeed one dimensional.

Using the electromagnetic acoustic transducer (EMAT) for low frequencies
that we have recently developed \cite{MoralesGutierrezFlores}, we
can excite and measure normal-mode frequencies and wave amplitudes
for torsional waves in metallic rods. The experimental apparatus has been
described in detail elsewhere \cite{MoralesFloresGutierrezMendez-Sanchez}.
For locally periodic systems a band structure emerges. As shown in
Ref. \cite{MoralesFloresGutierrezMendez-Sanchez}, these normal-mode
properties can also be computed using the transfer matrix method.
The theoretical results agree very well with our experimental measurements;
we should emphasize that this is a parameter free fit.

In Fig.~\ref{espectro} we present the band spectrum obtained for
$N=10$, both from the theoretical and experimental points of view. We see that
in the gap between the first and second bands only one frequency appears,
but in the second forbidden gap two levels lie. In some of the higher
gaps two frequencies are also found. This is an experimental realization of the
theoretical findings of Gumen et al \cite{Gumenetal}. In
Fig.~\ref{waveamplitudes} we show
the wave amplitudes of these two states; they are localized around
the topological defect with an exponential decay. The theoretical values,
both for the frequencies
and wave amplitudes coincide well with our measurements, as these
two figures show.

\begin{figure}
\vspace*{5mm}\includegraphics[width=\columnwidth]{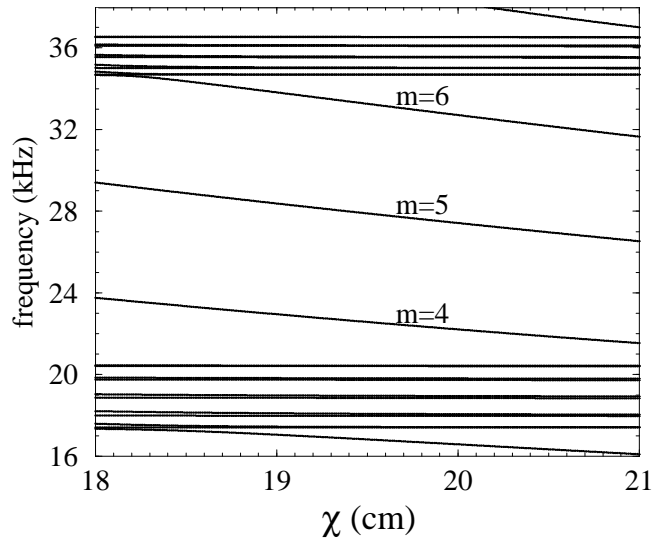}
\caption{Theoretical frequency spectrum as a function of $\chi$.
For $\chi=18$~cm {\it two} levels lie between the second and third bands.
For larger values of $\chi$ {\it three} levels are in the forbidden band.}
\label{chi}
\end{figure}

The theoretical results are easily extended for wider ranges of $\chi $.
The band spectrum as a function
of $\chi $ is given in Fig.~\ref{chi}.
Since the frequencies of the
localized states are proportional to $m/(l-\epsilon+\chi )$, where $m$ is
an integer number, more than one level can lie in the forbidden band.
For example, as will be seen in Fig.~\ref{chi}, for $\chi=18$~cm two levels
appear, whereas for higher values of $\chi$ three states are located in
the second forbidden band.

\section{CONCLUSION}

We have measured and calculated numerically normal-mode
frequencies and
amplitudes of torsional waves in a rod with a topological defect. The
theoretical prediction of Gumen {\it et al}
\cite{Gumenetal} that for certain topological defects two levels instead
of one become localized is shown to be true, and an experimental example
is provided. We also give, via a numerical calculation, a generalization
of the results found in Ref. \cite{Gumenetal}, since for other special
values of the topological defect more than two
frequencies can lie within the forbidden bands.

\begin{acknowledgments}
We would like to thank A. Krokhin for suggesting the problem. This
work was supported by DGAPA-UNAM Project No. IN104400.
\end{acknowledgments}


\begin{thebibliography}{1}
\bibitem{Gumenetal}L. Gumen, E. Feldman, V. Yurchenko, A. Krokhin, and P. Pereyra, Physica
E \emph{In Press.}
\bibitem{MoralesGutierrezFlores}A. Morales, L. Guti\'errez, and J. Flores
Am. J. Phys. {\bf 69}, (2001) 517.
\bibitem{MoralesFloresGutierrezMendez-Sanchez}A. Morales, J. Flores, L. Guti\'errez, and
R. M\'endez-S\'anchez, {\em J. Acoust. Soc. Am. \bf 112}, (2002) 1961 .
\end{thebibliography}
\end{document}